\documentclass[twocolumn,letterpaper]{IEEEAerospaceCLS}  
\usepackage[]{graphicx,float,latexsym,amssymb,amsfonts,amsmath,amstext,times}

\newcommand{\ignore}[1]{}  

\usepackage{url}
\usepackage[usenames,dvipsnames,svgnames,table]{xcolor}

\usepackage{stfloats}

\begin{document}

\title{Towards Space Solar Power - \\ Examining Atmospheric Interactions of Power Beams with the HAARP Facility}

\author{%
Martin Leitgab\\
Space Radiation Analysis Group, NASA JSC\\
Lockheed Martin Corporation \\
Houston, TX $77058$ \\
$+1-217-722-9353$ \\ 
Martin.Leitgab@nasa.gov
\and 
Aidan Cowley\\ 
School of Electronic Engineering\\
Dublin City University\\
Glasnevin, Dublin $9$, Co. Dublin, Ireland \\
$+353-1-700-7696$ \\ 
cowleya3@mail.dcu.ie
\thanks{\footnotesize 978-1-4799-1622-1/14/$\$31.00$ \copyright2014 IEEE.} 
}

\maketitle

\thispagestyle{plain}
\pagestyle{plain}

\begin{abstract}
In the most common space solar power (SSP) system architectures, solar energy harvested by large satellites in geostationary orbit is transmitted to Earth via microwave radiation. Currently, only limited information about the interactions of microwave beams with energy densities of several tens to hundreds of W/m$^2$ with the different layers of the atmosphere is available. Governmental bodies will likely require detailed investigations of safety and atmospheric effects of microwave power beams before issuing launch licenses for SSP satellite systems. This paper proposes to collect representative and comprehensive data of the interaction of power beams with the atmosphere by extending the infrastructure of the High Frequency Active Auroral Research Program (HAARP) facility in Alaska, USA. Estimates of the transmission infrastructure performance as well as measurement devices and scientific capabilities of possible upgrade scenarios will be discussed. The proposed upgrade of the HAARP facility is expected to deliver a wealth of data and information which could serve as a decision base for governmental launch licensing of SSP satellites, and which can be used in addition to deepen public acceptance of SSP as a large-scale renewable energy source.
\end{abstract}

\tableofcontents 


\section{Introduction}

Space solar power (SSP) is a promising candidate for large-scale renewable energy sources operable by the second half of this century. Many SSP system designs consist of large photovoltaic modules in geostationary orbit which transform sunlight to electricity. The generated power is used to drive microwave generation devices, which feed a phased array of microwave transmitters forming a collimated beam directed towards earth. While a nominal atmospheric attenuation of about $2\%$ for a microwave frequency of $2.45$~Ghz~\cite{ArndtNASAtechreport} is comparably small, yet $100$~MW of power is absorbed by the atmosphere in the case of a $5$~GW power station in geostationary orbit (GEO). Even though substantial amounts of energy are expected to be deposited in the atmosphere by utility-scale SSP satellites, only little information is available about atmospheric interactions of microwave power beams. Currently existing data was obtained from several sounding rocket and satellite experiments mostly conducted by Japanese scientists (for a review of previous measurements, see Ref.~\cite{ursipaper}). Due to limitations in transmitter size and transmissible power, the range of microwave beam energy densities comparable to beams from utility-scale SSP satellites is limited to the immediate atmospheric vicinity of the spacecraft~\cite{Tanakaiacpaper}. 

A solid base of in-depth knowledge and comprehensive studies documenting safety and interactions of power beams with all relevant layers of the atmosphere will likely be required to obtain launch and operating licenses for SSP systems from US national and international regulatory bodies such as the International Telecommunication Union. Such data could also largely facilitate the approval process of SSP launches in the context of the International Traffic in Arms Regulations (ITAR)~\cite{ITARweb}. In addition, scientific data of atmospheric interactions of power beams could, if properly published and disseminated, build public awareness and support and ease public concerns of possible hazards of SSP as a new renewable energy option. A favorable attitude of the general public towards SSP would likely facilitate the development of political and funding support for timely SSP research and development, prototyping and deployment. Aside from SSP-related benefits, the obtained data can also be used as valuable input to further inform and verify computational models of the atmosphere and to provide better understanding of its interaction with electromagnetic waves in general.

As a potentially more cost-effective and efficient alternative to on-orbit/high-altitude microwave power beam sources, the present paper proposes to utilize ground-based facilities to collect the large amounts of information needed to enable the deployment of a SSP demonstrator and, later on, utility-scale SSP satellites. Ground-based systems do not experience as stringent limitations in transmitter size and transmitted RF power levels as high-altitude microwave sources and offer continuous experimental accessibility. In addition, power beams generated at ground facilities can readily access the lowest levels of the atmosphere which likely benefits public acceptance of the safety of wireless power beaming. On the other hand, ground facilities can probe atmospheric layers up to very high altitudes if equipped with sufficient transmissible power, as will be shown below. Lastly, large amounts of data can be gathered over essentially unlimited operating times, while total cost between ground and orbit-based systems could be comparable if existing ground facilities are extended.

The study presented in this paper is intended to provide a first look at the option of using a ground-based facility to simulate and study interactions of SSP microwave power beams with the atmosphere. Such a project could become the first large-scale, concerted effort to establish comprehensive scientific knowledge of atmospheric power beam interactions and represent a critical impetus to start large-scale SSP development for a potentially modest amount of required funding, as described below. 


\section{Potential Effects of Microwave Beams in Atmosphere}

\begin{figure*}
 \centering
 \includegraphics[width=0.9\linewidth]{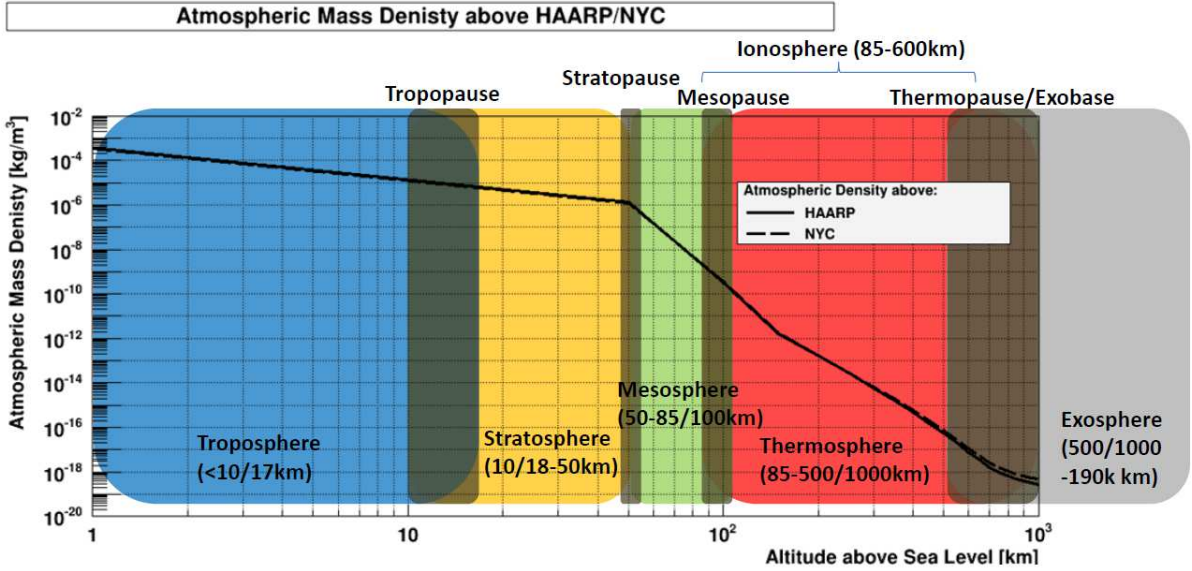}
 \caption{Atmospheric density versus altitude above sea level according to the NRL MSIS-$00$ model, above the HAARP facility and New York City, both USA. Relevant atmospheric layers are illustrated.}
 \label{fig:atmosmassdens}
\end{figure*}

To introduce notation used in the following and as an illustration, Figure~\ref{fig:atmosmassdens} shows the mass density of the Earth's atmosphere versus altitude above sea level according to the US Naval Research Laboratory (NRL) MSIS-$00$ model~\cite{nrlatmosdensmodel} above the High Frequency Active Auroral Research Program (HAARP) facility in Alaska and New York City, both USA, with an overlay of relevant atmospheric layers~\cite{atmoswikipedia}. Altitude ranges for atmospheric layers are intended for illustration only. 

The International Union of Radio Science (URSI) published a White Paper in $2007$ about several aspects of SSP~\cite{ursipaper}. This report lists several possible interactions of $2.45$ and $5.8$~GHz microwave beams with the different layers of the atmosphere and possible sources of interference with existing radio services due to the presence of microwave power beams, which will be described in the following. 

\subsection{Atmospheric Interactions of Microwave Power Beams}

Under a nominal space weather paradigm (i.e. in the absence of uncommon solar activity) it has been shown that most of the absorption of the power beam would occur in the lower ionosphere regions~\cite{a1} below $200$~km. The coherence and amplitude of the power beam is dependent on linear processes such as absorption due to electron collisional events, diffraction owing to plasma density variations and refraction. Of more interest are the generation of nonlinear interactions of the power beam and the atmosphere. Such nonlinear effects are notable in their difficulty to accurately model and predict. 

Non-linear interactions between microwave power beams and the atmosphere lead to a reduction in power transmission efficiency via additional absorption and heating of electrons. Electron or Ohmic heating has a direct impact of the atmospheric chemistry makeup as well as the electron density (and thus further collisional and scattering events). As the electron density increases as a function of beam power density, the thermal electron gas pressure increases. In the inhomogeneous gas environment of the various atmospheric altitudes, this will result in gas density gradients where the plasma is pushed out towards lower pressure regions resulting in localized electron density depletions. At altitudes below $200$~km, this change in the make up of the plasma volume leads to refractive effects, such as self-focusing, and also to localized regions where chemical reaction chains can occur owing to changes in $N_{e}$ and $T_{e}$, the electron density and temperature respectively.

In addition, Ref.~\cite{ursipaper} describes the potential for excitation of plasma waves in the ionosphere (including the meso-, thermo- and exosphere) by microwave power beams, which was experimentally observed by Japanese sounding rocket experiments. Such waves are considered to be a possible source of secondary electromagnetic waves which could generate additional Ohmic heating of the atmosphere or produce interference with radio services. Studies referenced in Ref.~\cite{ursipaper} describe a possible significant increase of the temperature in the thermopause and ozone creation due to microwave beams with large power densities of about $500$~W/m$^{2}$, which are however most likely beyond the maximum power densities achieved by microwave solar power satellites (SPS). Reference~\cite{ursipaper} also points out potential defocusing and amplitude fluctuations of the power beams caused by electron-density irregularities which occur naturally in the atmosphere and which could interfere with beam control. Lastly, Ref.~\cite{ursipaper} recommends studying the power beam refraction, attenuation and diffusion effects in the troposphere due to the presence of atmospheric gases and aerosols, or water/ice clouds and precipitation. In general, Ref.~\cite{ursipaper} states that non-linear and feedback effects can only be predicted with limited accuracy and have to be studied experimentally. 

\subsection{Interference of Microwave Power Beams with Radio Services}

In addition to atmospheric interactions of power beams, Ref.~\cite{ursipaper} points out that interference of power beams with existing radio services have to be examined in order for SSP to be able to legally operate under ITU regulations. Due to closeness in frequency, the $2.45$ and $5.8$~GHz bands allocated for civilian and military wireless applications and ISM (industry, science and medical) applications are expected to be at risk for interference, mainly from spurious or out-of-band emissions of SSP microwave transmission elements. Reference~\cite{ursipaper} also estimates possible interferences of microwave beams with high-sensitivity radio-astronomy detectors in the $4.9$ to $5.0$~GHz range, and with passive sensing devices in the $1.4$~GHz band used to investigate soil moisture and ocean salinity. Finally, Ref.~\cite{ursipaper} recommends to investigate all possible effects of power beams on the electrical conductivity and the chemical composition of atmospheric layers, even if not yet observed.

All of the above aspects of recommended and required investigation are accessible with the proposed extensions of the HAARP facility infrastructure discussed below, which therefore would represent a crucial step towards regulatory clearance and licensing of SSP infrastructure.


\section{The High Frequency Active Auroral Research Program (HAARP) Facility}
\label{sec:haarpdescr}
The HAARP facility~\cite{haarpwebsite} near Glennallen, Alaska, USA was built beginning in $1994$ by the US Air Force, Navy and Defense Advanced Research Projects Agency (DARPA). Its initial purpose was to investigate interactions of electromagnetic waves with the ionosphere to improve communication and surveillance systems, and to conduct additional scientific studies of the atmosphere. The facility houses a phased antenna array of size $300$~m $\times$ $370$~m to locally and temporarily excite areas in the ionosphere with $2.8$ to $10$~MHz electromagnetic waves at a maximum level of $3.6$~MW transmissible power. 

A number of on- and off-site scientific instruments is used to investigate the physical properties of the excited ionospheric areas. Among these, active instruments include the high frequency (HF) ionosonde to probe the electron density in the ionosphere and the ultra-high frequency (UHF) ionosphere radar investigating the propagation of radio wave signals in the ionosphere. Passive devices used in HAARP research are magnetometers to study variations of the geomagnetic field, relative ionosphere opacity meters ('riometers') measuring ionospheric radiation absorption, radio spectrometers for frequencies between $100$~kHz and $1$~GHz to monitor HAARP emissions, optical imagers/photometers, low frequency (LF) and ionospheric scintillation receivers such as total electron content receivers. The facility also houses an aircraft radar to avoid interference with flight traffic. 


\section{Using the HAARP Facility to Investigate Atmospheric Microwave Power Beam Interactions}

\subsection{Extension Scenarios of the HAARP Phased Transmission Array}
In order for the HAARP transmission array to produce low-GHz frequency microwave beams as used in common SSP architectures, additional dipole antennas of appropriate size (e.g. $12.2$~cm to achieve frequencies of $2.45$~GHz) have to be mounted on the existing transmission antenna masts of the HAARP transmission array. In addition, appropriate radiation generation devices and amplifiers (e.g. oscillators and solid state power amplifiers) have to be integrated with the existing HAARP infrastructure. To evaluate the capabilities of a microwave-modified HAARP array in emulating utility-scale SPS microwave power beams, two sets of criteria are formulated in the following concerning beam width and peak power density. For convenience, all power beam calculations are performed for $2.45$~GHz beam frequencies.

\subsubsection{Beam Size of HAARP Phased Array Upgrade Scenarios}
\label{sec:beamwidth}
Power beam diameters generated at HAARP will remain smaller than power beams from LEO SPS demonstrators and GEO utility-scale SPS plants up to high altitudes above the surface. However, it is estimated that all atmospheric conditions caused by microwave power beams requiring further study already occur on scales much smaller than the multi-km width of LEO or GEO beams on the ground. The present study assumes that qualitatively identical atmospheric conditions can be created by HAARP microwave beams as soon as the beam width is larger than $10$ to $100$~m. 

\begin{figure*}[t]
 \centering
 \includegraphics[width=0.9\linewidth]{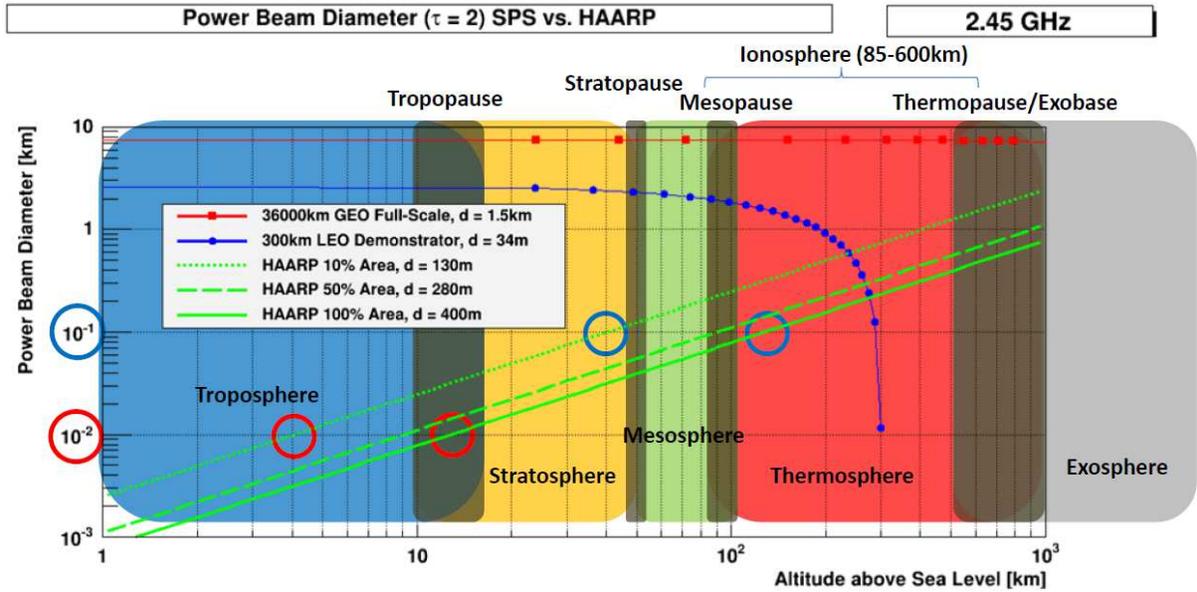}
 \caption{Diameter of power beams containing $95\%$ of transmitted power for HAARP microwave-modifications of $10\%$, $50\%$ and $100\%$ of the current HAARP transmitter array are shown, achieving relevant scales of $10$ to $100$~m for investigations of atmospheric interactions between $4$ and $130$~km altitude (indicated by red and blue circles, respectively). For illustration, power beam widths of an exemplary LEO demonstrator and GEO utility-scale SPS are shown as well.}
 \label{fig:beamdiam}
\end{figure*}

Three options for the proposed HAARP microwave-modification are evaluated, where $10\%$, $50\%$ and $100\%$ of the current HAARP transmission array's size are equipped with microwave dipole antennas. The beam width for HAARP power beams with optimum phased array focusing is calculated by the expression

\begin{equation}
\tau = \frac{\sqrt{A_{t}A_{b}}}{\lambda L},
\label{eq:taueq}
\end{equation}

with the choice $\tau = 2$ such that more than $95\%$ of the power transmitted from an aperture with area $A_{t}$ is contained within the beam area $A_{b}$ at altitude above sea level $L$~\cite{jaffepaper}, for transmitted wavelength $\lambda$. Figure~\ref{fig:beamdiam} illustrates the described comparisons. For illustration purposes, power beam diameters from a LEO demonstrator of about $34$~m diameter at an altitude of $300$~km proposed in Ref.~\cite{ownIACpaper} and from a GEO utility-scale power plant of about $1.5$~km diameter as described in Ref.~\cite{ownBaltimorepaper} are given as well. All transmission surfaces are assumed to be of circular shape.

It can be seen that for HAARP microwave-modifications effective on $10\%$ and $100\%$ of the current HAARP transmitter area, beam widths larger than $10$~m can be achieved for all altitudes higher than about $4$ and $14$~km (red circles), respectively, while beam diameters larger than $100$~m are obtained for altitudes larger than about $40$ and $130$~km (blue circles). The results shown in Figure~\ref{fig:beamdiam} indicate that all HAARP microwave-modification options would produce power beams sufficient in size to study atmospheric interactions with optimum transmitter array focusing for altitudes above $4$ to $130$~km. Nevertheless, also lower altitudes are readily accessible for study by appropriately defocussing the HAARP power beams. It is concluded that power beams of sufficient diameter to study potential large-scale atmospheric changes induced by LEO and GEO SPS power beams can be produced by an upgraded HAARP phased array infrastructure for all altitudes.  

\subsubsection{Power Density of HAARP Phased Array Upgrade Scenarios}
\label{sec:powerdens}
In order to representatively study atmospheric changes created by LEO demonstrator and GEO utility-scale SPS plants, power beams generated at the HAARP facility have to reach comparable energy densities. It is expected that qualitatively identical atmospheric effects can be generated by HAARP power beams compared to LEO and GEO SPS, as long as the HAARP peak power densities do not fall by more than a factor of $10$ below the SPS densities at the same altitude.

\begin{figure*}[t]
 \centering
 \includegraphics[width=0.9\linewidth]{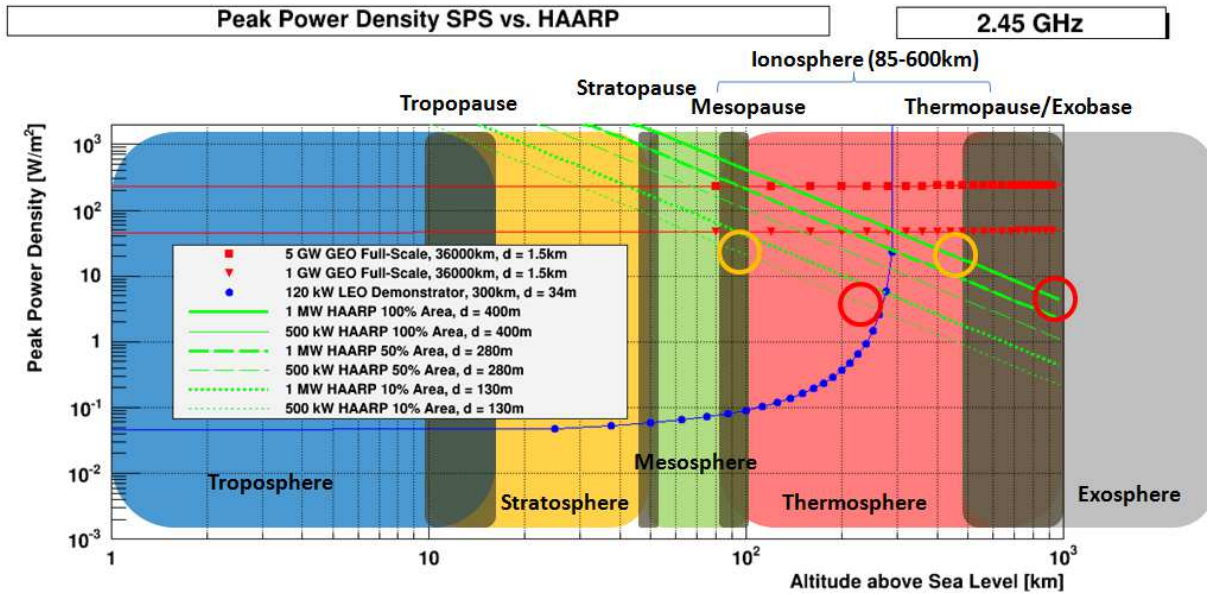}
 \caption{Peak power densities in the center of a Gaussian beam taper for HAARP microwave-modifications of $10\%$, $50\%$ and $100\%$ of the current HAARP transmitter array, for $500$~kW and $1$~MW transmitted power are shown. Relevant power densities for the introduction of power beam atmospheric interaction effects comparable to power beams from $1$~GW GEO SPS are achieved up to maximum altitudes of $210$~km and $930$~km for upgrade scenario (a) and (b), respectively (indicated by red circles). Maximum conceivable SPS power densities can be emulated up $95$~km and $420$~km for upgrade scenarios (a) and (b), respectively (orange circles). LEO demonstrator SPS power densities are surpassed by all upgrade options up to about $300$~km.}
 \label{fig:peakpowerdens}
\end{figure*}

Two upgrade options are considered in this Section, where $500$~kW and $1$~MW of HAARP's maximum transmissible power of $3.6$~MW are used to power the phased antenna array. Peak power density against orbital altitude is calculated by using an expression adapted from Ref.~\cite{iaastudy},

\begin{equation}
I_{\textnormal{peak}} = \pi \times \frac{P_{t}}{8} \times ( \frac{D_{t}}{\lambda L} )^{2},
\label{eq:powerdenseq}
\end{equation}

where $I_{\textnormal{peak}}$ represents the peak beam power density in the center of a Gaussian beam taper, $P_{t}$ the transmitted power, $D_{t}$ the transmitter diameter, $L$ the orbital altitude and $\lambda$ the transmission wavelength. Figure~\ref{fig:peakpowerdens} shows power densities for the three different upgrade options of HAARP phased array area from the previous Section in conjunction with the two upgrade options of HAARP RF power levels used. As a reference, peak power densities of an exemplary $120$~kW RF LEO demonstrator~\cite{ownIACpaper} and a $1$~GW RF, $1.5$~km diameter GEO utility-scale power plant~\cite{ownBaltimorepaper} are included in the plot as well. An upper limit of power densities from mid-term GEO microwave-based SPS is assumed to be equivalent to the power densities of a $5$~GW RF GEO SPS with $1.5$~km transmission diameter, which are also given in the plot.   

For the least-effort HAARP upgrade option involving $10\%$ of HAARP's phased array area and $500$~kW RF power, in the following referred to as upgrade scenario (a), Figure~\ref{fig:peakpowerdens} indicates that HAARP peak power densities remain within a factor of $10$ of the $1$~GW GEO full-scale plant up to altitudes of $210$~km (left red circle). If the entire HAARP phased array is upgraded to $2.45$~GHz and powered by $1$~MW, which will be referred to as upgrade scenario (b), peak beam power densities are comparable to power densities from the $1$~GW GEO power station beams within a factor of $10$ up to altitudes of about $930$~km (right red circle). 

For all upgrade options, LEO SPS demonstrator power densities are surpassed by HAARP peak power densities up to LEO orbits of about $300$~km. Even for the upper limit case of power beams from a $1.5$~km diameter, $5$~GW GEO SPS, representative data can be collected for altitudes through the stratosphere and mesosphere up to $95$~km in upgrade scenario (a) and up through half of the thermosphere to $420$~km altitude for the full upgrade scenario (b) (orange circles). 

It is summarized that while the smallest HAARP upgrade option already allows the collection of representative power beam atmospheric interaction data well into the thermosphere, the full upgrade makes all atmospheric layers up to the exosphere accessible to HAARP power beam studies. It shall be noted that additionally, power densities from HAARP beams are very large and reach beyond the maximum conceivable SPS energy densities at altitudes below about $30$~km for upgrade scenario (a) to $110$~km for upgrade scenario (b). Therefore, defocussed HAARP power beams achieving widths of $10$ to $100$~m also at low altitudes are nevertheless expected to attain representative peak power densities to study lower-altitude atmospheric interactions. 

As additional benefits, an upgraded HAARP facility could enable validation of several aspects of wireless power transmission. According to Figure~\ref{fig:peakpowerdens}, under upgrade scenario (b) a low-technology $10$~m $\times$ $10$~m rectenna grid satellite deployed in a $300$~km orbit could intercept about $5$~kW of HAARP RF power to demonstrate wireless power transmission concepts over long distances between objects on orbit and on the ground. Lastly, an upgraded HAARP array could serve as the first ground test rectenna for a SSP LEO demonstrator or a full-scale GEO SSP plant.  

\subsubsection{Cost Estimates of HAARP Extensions}

An order-of-magnitude cost estimate is attempted in this Section. Possible HAARP array extension scenarios will employ similar hardware as used in the construction of SSP rectennas. Therefore, an area-specific cost of $10$~USD/m$^{2}$ as quoted in previous SSP architecture studies (e.g.~\cite{ownBaltimorepaper}) is assumed to hold for HAARP upgrades as well. Since most of civil engineering and power infrastructure is already present at HAARP, this area-specific cost could be regarded as an upper limit for HAARP upgrade hardware cost. For simplicity, only the cost for hardware in the power beam generation and distribution on the phased array is considered. Cost for potential additional diagnostic equipment and operations are not included. Estimated cost levels for the different HAARP phased array upgrade scenarios are given in Table~\ref{tab:costest}.

\begin{table}
 \renewcommand{\arraystretch}{1.3}
 \caption{\bf Order-of-magnitude cost estimates for HAARP upgrade scenarios based on upgraded area of phased array and $10$~USD/m$^{2}$.}
 \label{tab:costest}
 \centering
 \begin{tabular}{|c|c|}
  \hline
  \bfseries Phased Array Upgrade & \bfseries Est. Cost \\
  \hline\hline
Upgrade $10\%$ (D=130~m)& 130k~USD \\ \hline
Upgrade $50\%$ (D=280~m)& 620k~USD \\ \hline
Upgrade $100\%$ (D=400~m)& 1.3M~USD \\ \hline
  \hline
 \end{tabular}
\end{table}

Considering the results of the previous paragraphs discussing HAARP and SPS power beam width and power density, an optimization of cost effectiveness could be achieved by pursuing a staged implementation process of the different upgrade scenarios. Initial cost can be kept low and in reach of obtainable funding with implementing upgrade scenario (a) on $10\%$ of the HAARP transmitter area and using $500$~kW RF power, which could provide the required fact base to license a LEO demonstrator with sizing and power scales as estimated above. The HAARP facility could be further upgraded during the following $5$ to $10$ years as development of GW-scale GEO SPS platforms progresses, such that data taken with the full upgrade scenario (b) can enable final licensing and deployment of full-scale GEO SPS. Opportunities for further study are given by a more detailed cost estimate for the required hardware upgrade of the power beam generation and distribution across the HAARP phased array for $2.45$~GHz, and an extension of the analysis to $5.8$~GHz of frequency.  

\subsection{Extension of Diagnostic Capabilities for Atmospheric Interactions of Power Beams}

As mentioned previously, the HAARP facility is already equipped with a number of on- and off-site facilities that are well suited for gathering experimental data on atmospheric interactions of a simulated SPS power beam. Especially, riometry systems are already available at the HAARP facility and can be augmented or expanded upon to provide the correct metrology for the proposed distinct $2.45$~GHz frequency. In addition, off-site complimentary experimental methodologies can be applied to maximize data collection with an upgraded HAARP facility. 

\subsubsection{Sounding Rocketry}
Similar to the previously discussed Japanese sounding rocket experiments, which provided some insight into the potential operation of a SPS power beam, sub-orbital test flights can provide useful information about atmospheric behavior. Spectrophotometric measurements of the helium, oxygen and hydrogen spectral lines were taken on the Black Brant II sounding rocket launch in $1988$~\cite{a2}, in order to characterize aspects of the thermosphere/exosphere plasma environment. Such spectral measurements could potentially be used to determine gas densities within targeted volumes, using the principle of chemical actinometry~\cite{a3}. Carrying out SPS power beam experiments with the modified HAARP array could be timed to coincide with a launch for experimental measurement of the plasma state and gas species densities within a targeted volume coincidental with the rocket flight trajectory.

\subsubsection{Weather Balloons}
Low-cost atmospheric soundings utilizing weather balloons as experimental platforms are widely deployed. Gas density and plasma state observations can be made using UV photometry, mass spectroscopy, and electrochemical sondes. Weather balloon experiments custom-designed for sensitivity to specific atmospheric parameters can be deployed around the HAARP facility site during a radiation campaign to collect additional data on atmospheric interactions of HAARP power beams.

\subsubsection{Earth Observation Satellites}
The HAARP Diagnostic Satellite Scintillation (SATSIN) system has been used previously to characterize the structure and dynamics of ionospheric irregularities created during HF heating, utilizing the perturbation and modulation of UHF/VHF signals from satellite beacons as they pass through the power beam volumetric area and near vicinity. This system offers additional diagnostic capabilities for the upgrade proposed in this paper as well by measuring ionosphere behavior (e.g. Ohmic heating response, $N_{e}$ and $T_{e}$ variations, observation of nonlinear effects).

In related work, a joint HAARP/NASA experiment to detect the effects of a high power radiowave transmission on the magnetosphere was carried out by the WIND satellite radio and plasma wave detectors (WAVE)~\cite{a4}. While the target was in the high altitude magnetospheric band, such future experiments could be tailored to be analogous to an SPS power beam and such measurements could prove insightful, especially for geostationary positioned SPS systems.


\section{Summary}

Only limited amounts of data are available concerning the interaction of microwave power beams with Earth's atmosphere. It is expected that in-depth knowledge of atmospheric power beam interactions is required for the certification and licensing process of SSP and for raising public acceptance of SSP. Review publications of microwave interactions with atmospheric layers suggest that further studies need to be conducted, and that these studies should be of experimental nature due to expected non-linear effects which are difficult to simulate. The data generated from these experimental studies would provide key metrics in terms of characterizing atmospheric interactions of SPS power beams. 

This paper presents a first analysis of a possible usage of the existing infrastructure of the HAARP facility in Alaska, USA towards establishing a large-scale and comprehensive knowledge base of scientific rigor about the interactions of power beams with all relevant atmospheric layers. The power beam emulation capabilities of different scales of extension scenarios are compared. The above studies suggest that already the smallest considered upgrade allows the collection of representative data describing the interactions of a LEO demonstrator SPS power beam with the atmosphere for modest cost. Additionally, atmospheric power beam interactions of utility-scale SPS can potentially be emulated in a representative way by the largest considered upgrade. 

In-depth knowledge of the interactions of high-energy-density microwave power beams with Earth's atmosphere is expected to be of critical importance to enable SPS deployment and large-scale development supported by US national government agencies, international entities as well as the general public. Utilizing existing ground-based facilities could represent a cost-effective and efficient way of gathering the required data for modest amounts of funding.


\section{Disclaimer}

Lockheed Martin Corporation does not endorse the presented content in any form.

\bibliographystyle{IEEEtran}

\thebiography

\begin{biographywithpic}
{Martin Leitgab}{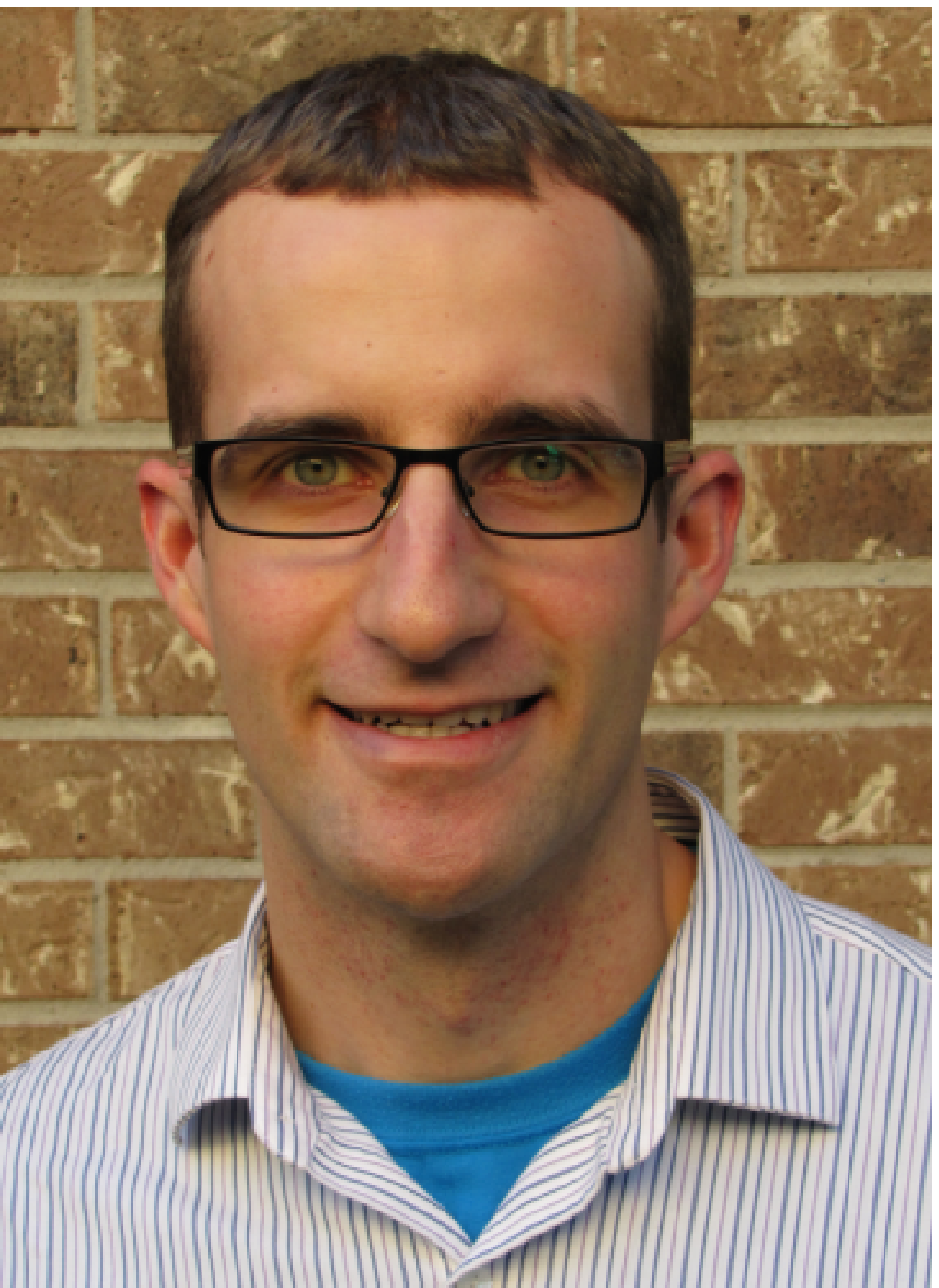}
received his 'Magister' degree (B.Sc./M.Sc. equivalent) in Physics from the University of Vienna, Austria in $2007$ and a Ph.D in Experimental Nuclear Physics from the University of Illinois at Urbana-Champaign, USA in $2013$. He is a senior research engineer at Lockheed Martin Corporation, working with the Space Radiation Analysis Group at NASA Johnson Space Center in Houston, TX, USA. He investigates neutron spectra on the International Space Station and currently focuses his SSP research efforts on the formulation of technology demonstration projects with the goal of enabling a SSP demonstrator mission in the near-term.
\end{biographywithpic} 

\begin{biographywithpic}
{Aidan Cowley}{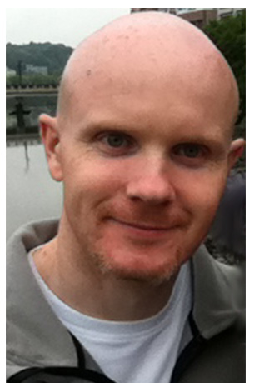}
received his B.Sc in Computer Applications and M.Eng in Electronic Systems from Dublin City University, Ireland, in 2004 and 2005 respectively, and a Ph.D in 2011. He is currently a researcher and lecturer at the National Centre for Plasma Science and Technology, Dublin City University, Ireland. His current research activities include novel optoelectronic materials, thermoelectrics, plasma deposition and metrology as well as renewable energy systems.
\end{biographywithpic} 


\end{document}